\documentclass[a4paper]{article}

\usepackage{INTERSPEECH2020}
\usepackage{multirow}

\title{A Joint Framework for Audio Tagging and Weakly Supervised Acoustic Event Detection Using DenseNet with Global Average Pooling}
\name{Chieh-Chi Kao\textsuperscript{1}, Bowen Shi\textsuperscript{2}, Ming Sun\textsuperscript{1}, Chao Wang\textsuperscript{1}}
\address{\textsuperscript{1}Alexa Speech, Amazon.com Inc. \textsuperscript{2}Toyota Technological Institute at Chicago}
\email{chiehchi@amazon.com, bshi@ttic.edu, \{mingsun,wngcha\}@amazon.com}

\begin{document}

\maketitle
\begin{abstract}
This paper proposes a network architecture mainly designed for audio tagging, which can also be used for weakly supervised acoustic event detection (AED).
The proposed network consists of a modified DenseNet as the feature extractor, and a global average pooling (GAP) layer to predict frame-level labels at inference time.
This architecture is inspired by the work proposed by Zhou et al., a well-known framework using GAP to localize visual objects given image-level labels.
While most of the previous works on weakly supervised AED used recurrent layers with attention-based mechanism to localize acoustic events, the proposed network directly localizes events using the feature map extracted by DenseNet without any recurrent layers.
In the audio tagging task of DCASE 2017, our method significantly outperforms the state-of-the-art method in F1 score by 5.3\% on the dev set, and 6.0\% on the eval set in terms of absolute values.
For weakly supervised AED task in DCASE 2018, our model outperforms the state-of-the-art method in event-based F1 by 8.1\% on the dev set, and 0.5\% on the eval set in terms of absolute values, by using data augmentation and tri-training to leverage unlabeled data.
\end{abstract}

\section{Introduction}
\label{sec:intro}
Audio tagging is the task of detecting the occurrence of certain events based on acoustic signals.
Recent releases of public datasets~\cite{AudioSet,DCASE2017challenge,Fonseca2018_DCASE} significantly stimulate the research in this field.
Hershey et al.~\cite{CNN_Hershey_2017} did a benchmark of different convolutional neural network (CNN) architectures on audio tagging using AudioSet, which is a dataset consisting of over 2 million audio clips from YouTube and an ontology of 527 classes.
DCASE 2017 Task 4 subtask A~\cite{DCASE2017challenge} focuses on audio tagging for the application of smart cars.
The winner of this challenge used a gated CNN with learnable gated linear units (GLU) to replace the ReLU activation  after each convolutional layer~\cite{WL_AED_Xu_2018}.
Yan et al.~\cite{Yan_ICASSP2019} further improved the above-mentioned architecture by inserting a feature selection structure after each GLU to exploit channel relationships.

Besides classifying audio recordings into different classes, AED requires predicting the onset and offset time of sound events.
DCASE 2017 Task 2~\cite{DCASE2017challenge} provides datasets with strong labels for detecting rare sound events (baby crying, glass breaking, and gunshot) within synthesized 30-second clips.
Most of the state-of-the-art AED models are based on convolutional recurrent neural network (CRNN).
The winner of this challenge~\cite{Lim2017} used 1D CNN with 2 layers of long short term memory (LSTM) layers to generate the frame level prediction.
Kao et al.~\cite{Kao18} used region-based CRNN for AED, which does not require post-processing for converting the prediction from frame-level to event-level.
Shen et al.~\cite{Shen19} used a temporal and a frequential attention model to improve the performance of CRNN.
Zhang et al.~\cite{Zhang19} gathered information at multiple resolutions to generate a time-frequency attention mask, which tells the model where to focus along both time and frequency axis.

Training such AED models in a fully-supervised manner can be very costly since annotating strong labels (onset/offset time) is labor-intensive and time-consuming.
Weakly supervised AED (also called multiple instance learning) is an efficient way to train AED models without using strong labels.
It uses weak labels (utterance-level labels) to train a model, where the trained model is still able to predict strong labels (frame-level labels) at inference time.
DCASE 2017 Task 4 subtask B~\cite{DCASE2017challenge} provides datasets for weakly supervised AED in driving environments.
The winner of DCASE 2017 challenge used an ensemble of CNNs with various lengths of analysis windows for multiple input scaling~\cite{Lee2017a}. 
He et al.~\cite{He19} proposed a hierarchical pooling structure to improve the performance of CRNN.
The effect of different pooling/attention methods on AED and audio tagging also have been analyzed in previous works~\cite{Wang_18,Pooling_Wang19,Kao_2020}.
DCASE 2018 Task 4~\cite{DCASE2018challenge} further extends weakly supervised AED in domestic environments by incorporating in domain and out-of-domain unlabeled samples.
Lu~\cite{Lu_DCASE2018} proposed a mean-teacher model with context-gating CRNN to utilize unlabeled in-domain data.
Liu~\cite{Liu2018} used a tagging model with pre-set thresholds to mine unlabeled data with high confidence.

Although GAP layer has been used with VGG-based feature extractor for both tagging and localization~\cite{Kumar_2018,Kumar_2020}, our experimental results on DCASE 2017 Task 4 dataset show that DenseNet~\cite{DenseNet_CVPR_2017} works better as a feature extractor. 
On the other hand, DenseNet has been used in AED related tasks but not with GAP for both tagging and localization.
Zhe et al.~\cite{Zhe_2018} chunked the input into small segments, and fed each segment to DenseNet to generate frame-wise prediction for AED.
Jeong et al.~\cite{Jeong_2018} used DenseNet for audio tagging but not for localization.
This paper proposes a network architecture mainly designed for audio tagging, which can also be used for weakly supervised AED.
It consists of a modified DenseNet~\cite{DenseNet_CVPR_2017} as the feature extractor, and a global average pooling (GAP) layer to predict frame-level labels at inference time.
We tested our method on DCASE 2017 Task 4 subtask A for audio tagging, and the proposed method significantly outperforms the state-of-the-art method~\cite{Yan_ICASSP2019}.
We also tested our system for weakly supervised AED in driving environments (DCASE 2017 Task 4 subtask B) and domestic environments (DCASE 2018 Task 4).
Our method outperforms the state-of-the-art work~\cite{Dinkel2019} of DCASE 2018 Task 4 by using tri-training~\cite{TriTraining,TriTraining_Shi19} to leverage unlabeled data.

\begin{figure*}[t!]
    \centering
    \includegraphics[width=0.95\textwidth]{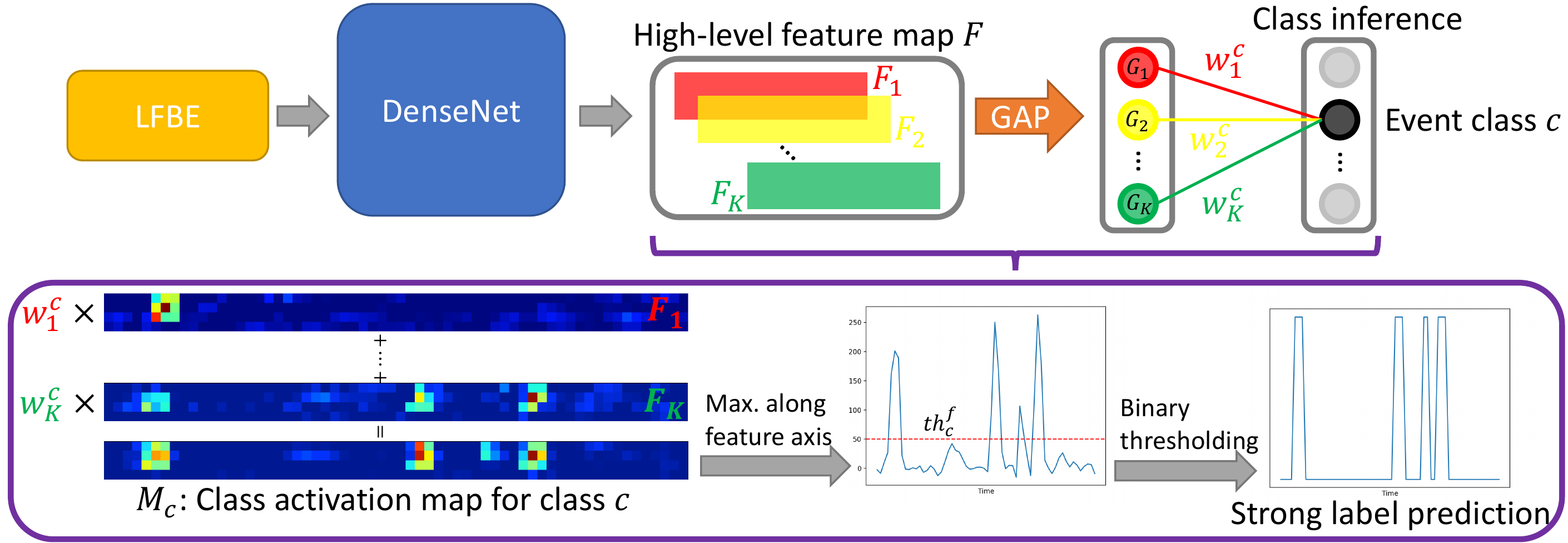}
    \caption{System overview of the proposed architecture for weakly supervised AED.}
    \label{fig:flowchart}
    \vspace{-0.2in}
\end{figure*}

\section{Proposed Method}
\label{sec:method}
The proposed network consists of a modified DenseNet~\cite{DenseNet_CVPR_2017} as a feature extractor, and a GAP layer for predicting frame-level labels at inference time.
In order to generate strong labels with finer resolution in time at inference, we modified DenseNet to have less pooling operations to maintain the resolution in time of the extracted feature map.
The exact network configurations we used are shown in Table~\ref{tab:arch}.
We used DenseNet-63 on DCASE 2017 Task 4 and DenseNet-120 on DCASE 2018 Task 4, and these architectures are chosen based on our experimental results on the dev set.

Given weak labels (i.e. utterance-level labels), the network can be trained under a multi-class classification setting.
Since multiple events of different classes can happen within the same utterance, we use sigmoid as the activation function with binary cross-entropy for each class.
We use the method proposed by Zhou et al.~\cite{GAP_CVPR_2016} to generate class activation maps (CAM) for predicting strong labels at inference time.
The system overview is shown in Fig.~\ref{fig:flowchart}.
Given an input utterance, a high-level feature map $F$ ($T \times N \times K$) can be extracted by DenseNet (i.e. input to the GAP layer), where $T$, $N$, $K$ represent the dimension in time, feature, and channel.
For each channel $k$, the GAP layer will generate a response $G_k$, which is the average of all features in channel $k$.
These responses are further fed into a dense layer to predict the classification probability.
For a given class $c$, the input to the sigmoid is $S_c=\sum_{k}w_{k}^{c}G_{k}$, where $w_{k}^{c}$ is the weight in the final dense layer corresponding to class $c$ for channel $k$.
The utterance-level prediction for class $c$ is $y_c=sigmoid(S_c)$.
$w_{k}^{c}$ controls the contribution of a given channel $k$ to class $c$.
The CAM for class $c$ is defined as:
\begin{equation}
M_{c} = \sum_{k}w_{k}^{c}F_{k},
\end{equation} 
where $F_{k}$ is channel $k$ of the high-level feature map $F$.

If one clip has utterance-level probability ($y_c$) greater than the utterance-level threshold ($th_{c}^u$, where $u$ represents utterance) at inference time, it indicates the occurrence of target class $c$. 
We can use CAM to predict strong labels.
We first convert the 2D CAM ($T \times N$) to a 1D sequential signal (length $T$) by taking the maximum value across the feature axis.
Strong labels of class $c$ are predicted by binary thresholding on the sequential signal with a frame-level threshold ($th_{c}^f$).
Note that the time resolution of the sequential signal is not the same as one frame in the input feature to the network (10 ms) due to pooling operations in the network.
Both utterance-level and frame-level thresholds are set by optimizing the F1 score of weakly supervised AED on the development set.

\section{Experimental Setups}
\label{sec:setups}
We tested our method on DCASE 2017 Task 4~\cite{DCASE2017challenge} and DCASE 2018 Task 4~\cite{DCASE2018challenge}.
Both of these two datasets are subsets of AudioSet~\cite{AudioSet}.
The audio clips are mono-channel and sampled at 44.1k Hz with a maximum duration of 10 seconds.
We decompose each clip into a sequence of 25 ms frames with a 10 ms shift.
64 dimensional log filter bank energies (LFBEs) are calculated for each frame, and we aggregate the LFBEs from all frames to generate the input spectrogram.
Note that we train all models in this work from scratch without any pre-training using external datasets, which is complied with task rules of DCASE Challenge.

\begin{table}
  \center
  \begin{tabular}{c|c|c}
    \hline
    Layers & DenseNet-63 & DenseNet-120 \\ 
     & (for DCASE2017) & (for DCASE2018) \\ \hline
    Convolution & \multicolumn{2}{c}{ 7 $\times$ 7 conv, stride 2} \\ \hline
    Dense Block (1) &  $\begin{bmatrix}  1 \times 1 \text{conv} \\ 3 \times 3 \text{conv}\end{bmatrix} \times 3 $& $\begin{bmatrix}  1 \times 1 \text{conv} \\ 3 \times 3 \text{conv}\end{bmatrix} \times 6 $\\ \hline
    Transition (1) & \multicolumn{2}{c}{1 $\times$ 1 conv} \\ 
    & \multicolumn{2}{c}{2 $\times$ 2 avg. pooling, stride 2} \\ \hline
  Dense Block (2) &  $\begin{bmatrix}  1 \times 1 \text{conv} \\ 3 \times 3 \text{conv}\end{bmatrix} \times 6 $& $\begin{bmatrix}  1 \times 1 \text{conv} \\ 3 \times 3 \text{conv}\end{bmatrix} \times 12 $\\ \hline
    Transition (2) & \multicolumn{2}{c}{1 $\times$ 1 conv} \\ 
    & \multicolumn{2}{c}{2 $\times$ 2 avg. pooling, stride 2} \\ \hline
    Dense Block (3) &  $\begin{bmatrix}  1 \times 1 \text{conv} \\ 3 \times 3 \text{conv}\end{bmatrix} \times 12 $& $\begin{bmatrix}  1 \times 1 \text{conv} \\ 3 \times 3 \text{conv}\end{bmatrix} \times 24 $\\ \hline
    Transition (3)  & 1 $\times$ 1 conv, 2 $\times$ 2  & N/A \\ 
    &avg. pool., stride 2& \\ \hline
  Dense Block (4) &  $\begin{bmatrix}  1 \times 1 \text{conv} \\ 3 \times 3 \text{conv}\end{bmatrix} \times 8 $& $\begin{bmatrix}  1 \times 1 \text{conv} \\ 3 \times 3 \text{conv}\end{bmatrix} \times 16 $\\ \hline
    GAP & \multicolumn{2}{c}{global avg. pooling} \\ \hline
    Classification & 17D dense, & 10D dense, \\ 
     & sigmoid & sigmoid \\ \hline

  \end{tabular}
  \caption{\label{tab:arch} DenseNet architectures for audio tagging and weakly supervised acoustic event detection. Note that each ``conv'' layer in dense blocks/ transition layers corresponds the sequence BN-ReLU-Conv. We set the growth rate to 32 as proposed in the original DenseNet~\cite{DenseNet_CVPR_2017}. Less pooling operations are used compared to the original DenseNet in order to have finer resolution in time.}
\end{table}

\subsection{DCASE 2017 Task 4} 
\label{sec:setups_2017}
There are two subtasks in this challenge: (A) audio tagging, (B) weakly supervised AED.
It contains 17 classes of warning and vehicle sounds related to driving environments.
The training set has only weak labels denoting the presence of events, and strong labels with timestamps are provided in dev/eval sets for evaluation. 
There are 51,172, 488, and 1,103 samples in train, dev, and eval sets, respectively.
We use the same metrics used in the challenge to evaluate our method.
For audio tagging, classification F1 score is used; for weakly supervised AED, we use segment-based F1 score~\cite{Mesaros2016_MDPI}, and the length of segments is set to 1 second.

We train DenseNet-63 model shown in Table~\ref{tab:arch} with adaptive momentum (ADAM) optimizer and the initial learning rate is set to 0.01.
The training is stopped when the classification F1 score on the dev set has stopped improving for 20 epochs.
We further finetune the model for 10 epochs with decreasing the learning rate to 0.001.
The size of minibatch is set to 200. 
For the results shown in the paper on DCASE 2017, we use an ensemble of 5 models by taking the average of output probabilities.
These 5 models are trained using the same hyper-parameters, and the only difference between them is the randomness in weight initialization and the data shuffling during training.

\subsection{DCASE 2018 Task 4} 
Task 4 of DCASE 2018 challenge consists of detecting onset/offset timestamps of sound events using audio with both weakly labeled data and unlabeled data. 
It contains 10 classes of audio events in domestic environments (e.g. Speech, Dog, Blender, etc.)
There are three different sets of training data provided: weakly labeled data, in-domain unlabeled data and out-of-domain unlabeled data.
Weakly labeled training set contains 1,578 clips with 2,244 occurrences with only utterance-level labels. 
The in-domain unlabeled training set contains 14,412 clips of which the distribution per class is close to the labeled set.
In addition, the unlabeled out-of-domain training set is composed of 39,999 clips from classes not considered in this task.
Note that event-based F1 is chosen by the challenge organizer as the evaluation metric, which is different from the segment-based F1 used in DCASE 2017 task 4B.

To utilize the unlabeled in-domain data, we use the tri-training proposed for audio tagging tasks in \cite{TriTraining_Shi19}. 
The idea of tri-training is similar to self-training, which takes advantage of a model trained with labeled data only to assign pseudo-labels to unlabeled data. 
Instead of relying on one model for pseudo-labeling, we train three independent models.
To update one of those three models, an unlabled clip gets a pseudo-label and is added into the training set if the other two models predict the same label with high confidence on the clip.
Generating pseudo-labels using consensus of multiple models mitigates mistakes made by a specific model.
One caveat of tri-training is that multiple models should differ such that the prediction of individual models complement each other. 
Although the training set is bootstrapped three times for training three models in \cite{TriTraining_Shi19}, we use the same training set while initializing models with different random seeds rather than bootstrapping. 
We find such practice leads to better performance which might be due to the limited amount of labeled data.

While predicting pseudo-labels of unlabeled data, we only infer utterance-level label. 
Model is trained with ADAM optimizer with an initial learning rate of 0.001 for 30 epochs, and the learning rate is reduced by half every 10 epochs. 
We chose the best weights out of 30 epochs based on classification F1 on the dev set. 
The batch size is set to 48 due to GPU memory constraints.
We also augment the labeled data by doing (1) circular shifting audio at a random timestep (2) randomly mixing two audio clips. 
When two clips are mixed, their labels are also merged. 
The number of labeled audio in augmented dataset is increased to 3,578. 
Only in-domain labeled data are used for pseudo-labeling in tri-training. 
For post-processing, we apply median filtering on the output segmentation mask, and the filter size per event is tuned based on event-based F1 on the dev set. 


\begin{table}
  \center
  \begin{tabular}{c|c|c}
    \hline
    Classification F1 & Dev (\%) & Eval (\%) \\ \hline
    Xu et al. \cite{WL_AED_Xu_2018} (ranked 1st) & 57.7 & 55.6 \\ \hline
    Lee et al. \cite{Lee2017a} (ranked 2nd) & 57.0 & 52.6 \\ \hline 
    Iqbal et al. \cite{iqbal2018capsule} & N/A & 58.6 \\ \hline 
    Wang et al. \cite{Pooling_Wang19} & 53.8 & N/A \\ \hline    
    Yan et al. \cite{Yan_ICASSP2019} & 59.5 & 60.1 \\ \hline    
    Ours & \textbf{64.8} & \textbf{66.1} \\ \hline    
  \end{tabular}
  \caption{\label{tab:DCASE2017_AT}Results on DCASE 2017 task 4A: audio tagging for smart cars}
\end{table}

\section{Experimental Results}
\subsection{Audio Tagging}
\label{sec:AT}
Table~\ref{tab:DCASE2017_AT} shows the classification F1 for the audio tagging subtask in DCASE 2017 task 4 on the development set and the evaluation set.
While most of the previous works of joint framework for audio tagging and weakly supervised AED use attention mechanism (e.g. gated CNN~\cite{WL_AED_Xu_2018}, attention by capsule routing~\cite{iqbal2018capsule}, region-based attention~\cite{Yan_ICASSP2019}, etc.), our method without any attention mechanism performs the best in audio tagging.
The proposed method outperforms the state-of-the-art method~\cite{Yan_ICASSP2019} in F1 score by 5.3\% on the dev set, and 6.0\% on the eval set.
Based on these results, we argue that attention mechanism may not be necessary for audio tagging.

\subsection{Weakly Supervised AED}
\label{sec:WL_AED}
\textbf{DCASE 2017:} Table~\ref{tab:DCASE2017_AED} shows the segment-based F1 for the weakly supervised AED subtask in DCASE 2017 task 4 on the development set and the evaluation set.
Although our method performs well on the audio tagging subtask, it does not outperform state-of-the-art methods in the weakly supervised AED subtask.
We suspect that the lack of attention mechanism may cause this performance gap in weakly supervised AED.
Exploring adding attention mechanism to our current model would be our future work.
We plan to explore whether it can improve the performance on weakly supervised AED, and how it impacts the performance on audio tagging.

\begin{table}
  \center
  \begin{tabular}{c|c|c}
    \hline
    Segment-based F1 & Dev (\%) & Eval (\%) \\ \hline
    Lee et al. \cite{Lee2017a} (ranked 1st) & 47.1 & \textbf{55.5} \\ \hline 
    Xu et al. \cite{WL_AED_Xu_2018} (ranked 2nd) & 49.7 & 51.8 \\ \hline
    Iqbal et al. \cite{iqbal2018capsule} & N/A & 46.3 \\ \hline 
    Wang et al. \cite{Pooling_Wang19} & 46.8 & N/A \\ \hline    
    Yan et al. \cite{Yan_ICASSP2019} & \textbf{51.3} & 55.1 \\ \hline    
    He et al. \cite{He19} & 46.5  & 53.4 \\ \hline    
    Ours & 49.9 & 49.4 \\ \hline    
  \end{tabular}
  \caption{\label{tab:DCASE2017_AED}Results on DCASE 2017 task 4B: weakly supervised AED for smart cars}
\end{table}

\begin{table}[t]
  \center
  \begin{tabular}{c|c|c}
    \hline
    Event-based F1 & Dev (\%) & Eval (\%) \\ \hline
    Lu et al. \cite{Lu_DCASE2018} (ranked 1st) & 25.9 & 32.4 \\ \hline
    Liu et al. \cite{Liu2018} (ranked 2nd) & \textbf{51.6} & 29.9 \\ \hline    
    Kong et al. \cite{KongBaseline2018} (ranked 3rd) & 26.7 & 24.0 \\ \hline
    Dinkel et al. \cite{Dinkel2019} & 36.4 & 32.5 \\ \hline
    Ours& 44.5 & \textbf{33.0} \\ \hline
  \end{tabular}
  \caption{\label{tab:DCASE2018}Results on DCASE 2018 task 4: weakly supervised AED in domestic environments}
\end{table}

\noindent\textbf{DCASE 2018:} We also tested our method on DCASE 2018 task 4, and the results are shown in Table~\ref{tab:DCASE2018}.
Different from the results on DCASE 2017 task 4, our method outperforms the state-of-the-art method~\cite{Dinkel2019} in event-based F1 by 8.1\% on the dev set, and 0.5\% on the eval set.
In order to know which part gives us the performance gain, we did an ablation study on this task.
As shown in Table~\ref{tab:DCASE2018_ablation}, data augmentation (cicular shifting and clip mixing) plays an important role, which might be due the amount of labeled training data is limited given model architecture is relative complicated. 
On top of that, using tri-training provides additional boost, which is complementary to data augmentation. 
For tri-training, we use an ensemble of six models, which consists of three models trained on labeled data only, and three models trained on both labeled data and in-domain unlabeled data.
If only labeled data are used, we use an ensemble of three models.
Note that the gap between dev and eval set, which is also observed in~\cite{Dinkel2019,KongBaseline2018,Liu2018}, might be due to the disparity of distribution of two sets.

\begin{table}[t]
  \center
  \begin{tabular}{c|c|c}
    \hline
    Event-based F1 & Dev (\%) & Eval (\%) \\ \hline
    Labeled data only & 34.9 & 25.8 \\ \hline
    $+$ data aug. & 42.0 & 29.5 \\ \hline
    $+$ data aug. \& unlabeled data& 44.5 & 33.0 \\ \hline
  \end{tabular}
  \caption{\label{tab:DCASE2018_ablation}Ablation study of data augmentation methods on DCASE 2018 task 4}
\end{table}

\section{Ablation Study}
\subsection{Feature extractor}
To investigate the performance of different feature extractors, we experimented with different architectures to generate the high-level feature map.
Three different types have been tested: VGG~\cite{VGG}, ResNet~\cite{ResNet}, and DenseNet~\cite{DenseNet_CVPR_2017}.
We modified each architecture to have similar number of parameters for fair comparison.
For VGG, the architecture is similar to the ConvNet configuration D in~\cite{VGG} with only 4 blocks and 9 conv layers.
For ResNet, the architecture is similar to ResNet-18 in~\cite{ResNet} with less number of filters in each block (from [64, 128, 256, 512] to [28, 56, 112, 224]).
For DenseNet, the architecture is described as DenseNet-63 in Table~\ref{tab:arch}.
The number of parameters of VGG, ResNet, DenseNet are 2.33M, 2.71M, and 2.34M.
Table~\ref{tab:DCASE2017_FE} shows the results on DCASE 2017 task 4 development set.
Note that all these results are based on ensemble of 5 models, which is the same setup as described in Sec.~\ref{sec:setups_2017}.
As shown in Table~\ref{tab:DCASE2017_FE}, DenseNet outperforms VGG and ResNet on both audio tagging (classification F1) and weakly-supervised AED (segment-based F1).
Based on these results, we chose DenseNet as the feature extractor through our experiments.

\begin{table}
  \center
  \begin{tabular}{c|c|c}
    \hline
     & Classification F1 (\%) & Segment-based F1 (\%) \\ \hline
    VGG & 63.5 & 48.9 \\ \hline
    ResNet & 62.4 & 48.9 \\ \hline  
    DenseNet & \textbf{64.8} & \textbf{49.9} \\ \hline    
  \end{tabular}
  \caption{\label{tab:DCASE2017_FE} Ablation study of different feature extractors on DCASE 2017 task 4 development set.}
\end{table}

\subsection{Class-wise performance for weakly-supervised AED}

To disentangle the effects of data augmentation and using unlabeled data, we did a further class-wise ablation study (see Table~\ref{tab:DCASE2018_eventwise}). 
Most events benefit from the both methods. 
As shown in Table~\ref{tab:DCASE2018_eventwise}, data augmentation helps detection of ``dishes'' and ``cat'' sound the most. 
We notice those events are generally short and are the foreground sounds in the original audio. 
Mixing audios provides richer background noise which helps the model disentangling the foreground sound from other sound. 
The gain brought by employing unlabeled data is related to the amount of labeled data, as we don't see large improvement from the ``speech'' event that has the largest amount of labeled data. 
Additionally, it is potentially related to the difficulty of detecting certain events. 
As some events are harder to detect (e.g., alarm/bell/ringing, running water) potentially due to the low loudness, ambiguity of definition and large variation, larger amount of training data are required to achieve high performance.
As a consequence, those events generally benefit more from the ways of increasing data amount including our semi-supervised approach and data augmentation.

\begin{table}
  \center
  \begin{tabular}{c|c|c|c|c}\hline
    \multirow{3}{*}{Event} &\multirow{3}{*}{\# clips} & \multicolumn{3}{c}{Evaluation F1 (\%)} \\\cline{3-5}
    & & label &  + data &  + data aug. \\ 
    & &  & aug. & \&unlabeled \\ \hline
    Dog & 214 & 13.0 &  17.3 &  \textbf{20.9} \\ \hline
Alarm/bell/ringing &  205 & 24.4 &  30.2 &  \textbf{37.5} \\ \hline
Speech  & 550 & 42.6 &  \textbf{44.7} & 44.4 \\ \hline
Blender & 134 & 13.4 &  18.7 &  \textbf{19.1} \\ \hline
Frying  & 171 & 42.7 &  \textbf{45.0} & 41.6 \\ \hline
Dishes  & 184 & 15.5 &  24.2 &  \textbf{26.6} \\ \hline
Running water & 343 & 15.0 &  17.0  & \textbf{25.0} \\ \hline
Cat & 173 & 9.5 & 17.7 &  \textbf{21.1} \\ \hline
Vacuum cleaner  & 167 & 37.4 &  33.6 &  \textbf{45.1} \\ \hline
Electric shaver & 103 & 44.2 &  46.4 &  \textbf{48.6} \\ \hline
  \end{tabular}
  \caption{\label{tab:DCASE2018_eventwise} Class-wise ablation study on DCASE 2018 task 4}
\end{table}

\section{Conclusions}
This paper proposes a network architecture mainly designed for audio tagging, which can also be used for weakly supervised AED.
Different from most of the previous works on weakly supervised AED that use recurrent layers with attention-based mechanism to localize acoustic events, the proposed network directly localizes events using the feature map extracted by DenseNet without any recurrent layers.
In the audio tagging task of DCASE 2017~\cite{DCASE2017challenge}, our method significantly outperforms the state-of-the-art method~\cite{Yan_ICASSP2019} by 5.3\% on the dev set, and 6.0\% on the eval set in F1 score.
For weakly supervised AED task in DCASE 2018~\cite{DCASE2018challenge}, our model outperforms the state-of-the-art method~\cite{Dinkel2019} by using data augmentaion and tri-training~\cite{TriTraining_Shi19} to leverage unlabled data.


\bibliographystyle{IEEEtran}
\bibliography{refs}


\end{document}